\documentclass[conference,a4paper]{APSIPA2023}
\usepackage{multirow}
\usepackage{amsmath}
\usepackage[psamsfonts]{amssymb}
\usepackage{amsxtra}
\usepackage{threeparttable}
\usepackage{graphicx}

\usepackage{geometry}
\usepackage{fancyhdr}

\fancypagestyle{firststyle} {
    \fancyhf{}
    \fancyhead[C]{2023 Asia Pacific Signal and Information Processing Association Annual Summit and Conference (APSIPA ASC)}

}

\begin{document}

\title{Ensembling Multilingual Pre-Trained Models for Predicting Multi-Label Regression Emotion Share from Speech}

\author{%
\authorblockN{%
Bagus Tris Atmaja\authorrefmark{1}$^{~~}$ and
Akira Sasou\authorrefmark{1}
}
\authorblockA{%
\authorrefmark{1}
National Institute of Advanced Industrial Science and Technology, Japan  \\
E-mail: b-atmaja@aist.go.jp, a-sasou@aist.go.jp} \\}
%


\maketitle
\thispagestyle{firststyle}
\pagestyle{fancy}

\begin{abstract}
Speech emotion recognition has evolved from research to practical applications. Previous studies of emotion recognition from speech have focused on developing models on certain datasets like IEMOCAP. The lack of data in the domain of emotion modeling emerges as a challenge to evaluate models in the other dataset, as well as to evaluate speech emotion recognition models that work in a multilingual setting. This paper proposes an ensemble learning to fuse results of pre-trained models for emotion share recognition from speech. The models were chosen to accommodate multilingual data from English and Spanish. The results show that ensemble learning can improve the performance of the baseline model with a single model and the previous best model from the late fusion. The performance is measured using the Spearman rank correlation coefficient since the task is a regression problem with ranking values. A Spearman rank correlation coefficient of 0.537 is reported for the test set, while for the development set, the score is 0.524. These scores are higher than the previous study of a fusion method from monolingual data, which achieved scores of 0.476 for the test and 0.470 for the development.
\end{abstract}


\section{Introduction}
Research on speech emotion recognition (SER) has been focused mainly on recognizing categorial emotions like happy, angry, sad, or neutral. There are attempts to predict dimensional emotions like valence, arousal, and dominance; however, their application is still limited. The reason is that the dimensional emotion is not as straightforward as the categorical emotion. Nevertheless, the previous studies on categorial emotion recognition are designed to choose a single class of emotions in multiclass settings. In contrast to categorial emotions, the emotions, particularly conveyed by vocal bursts, are bridged by smooth gradients (instead of clear separation between discreet emotions) with continuously varying meanings \cite{Cowen2019}. In that study, the authors found 24 categorial emotions, which may not have been limited to vocal bursts only but also to general emotion perception or speech perception (since vocal bursts are commonly found within speech). 

An attempt to analyze emotion fine-grained emotion (i.e., as a regression task) instead of coarse-grained emotion (i.e., by classification task) has been conducted from a linguistic point of view \cite{Buechel2016}. The authors mapped the degrees of valence and arousal given the sentence into six basic categorical emotions via several lexicons and corpus. The top model (using WKB lexicon) obtained correlation scores of 0.65 for dimensional emotions (valence, arousal, dominance) and 0.416 for categorial emotions. Besides only using linguistic information, the evaluation was conducted on English data only.

Recently, the multilingual approach to tackle the limitation of speech technology has been proposed in the parent tasks of speech processing, automatic speech recognition, speech synthesis, and speech translation. Models like Whisper \cite{Radford2021} and MMS \cite{Pratap} can precisely recognize speech in multiple languages. Stirred by this direction, research on speech emotion recognition has been directed to tackle its limitation under multilingual settings. Assuming that emotion is a universal language (i.e., the pattern of angry speech is the same across languages or cultures), the multilingual approach is expected to work also on speech-emotion recognition with diverse language samples.

Multilingual SER attempts to recognize emotion regardless of the language. One can build a single model (including a model trained on multilingual data) for each language with language detection as pre-processing; another can build a single model for all languages. This study is aimed at the latter approach. The previous studies utilized transfer learning for cross-lingual SER \cite{Upadhyay2023}, distribution adaption \cite{Li2023}, and implicit distribution alignment \cite{Zhao}. The last two studies were developed for cross-corpus SER with different languages. 

The use of multiple models for SER has shown to be effective for monolingual data. The technique of ensemble was usually adopted to fuse several models from different modalities and/or features. Examples are acoustic and linguistic for dimensional SER \cite{Atmaja2020h,Atmaja2020e}, acoustic and facial expression \cite{Zhao2018}, and multi-feature fusion \cite{Atmaja2020f}. In the previous study \cite{Shih2017}, ensemble learning was also utilized to tackle the limitation of skewed training data in monolingual SER by maximizing, averaging, and log averaging of the output probabilities of the component classifiers.

This paper contributes to the evaluation of ensemble learning, combining several models via late fusion methods for predicting emotion share from speech in multilingual settings. The fusion was performed by averaging predictions of different models. We evaluated eight multilingual pre-trained models since the dataset contains multilingual data from English and Spanish. The pre-trained models are chosen from those languages. We showed that by combining the models via ensemble learning with the late fusion method, we could improve the performance of the baseline model with a single model. The performance is measured using the Spearman rank correlation coefficient since the task is a regression problem with ranking values.

\section{Dataset and Task}

The dataset used in this study is obtained from the dataset for The ACM Multimedia 2023 Computational Paralinguistic Challenge (ComParE) Emotion Request Sub-Challange \cite{Schuller2022a}, which is provided by Hume AI \cite{Cowen2019a}. There are nine categorical emotions in the dataset: `Anger', `Boredom', `Calmness', `Concentration', `Determination', `Excitement', `Interest', `Sadness', and `Tiredness'. The choice of these nine emotions is based on their more balanced distribution in the valence-arousal space. The basis for the dataset itself is more than 5,000 `seed' samples. 

Seeds consist of various emotional expressions, which were gathered from openly available datasets, including MELD \cite{Poria2019} and VENEC \cite{Laukka2010}. The seed samples were mimicked by speakers recruited via Amazon Mechanical Turk by the provider of the dataset \cite{Cowen2019a}. The dataset consists of 51,881 `mimic' samples (total of 41:48:55 h of data, mean 2.9 s, range 1.2 - 7.98 s) from 1,004 speakers aged from 20 to 66 years old. The data were collected from 3 countries with broadly differing cultures: the United States (English/EN), South Africa (English/EN), and Venezuela (Spanish/SP). Each seed sample was rated by the individual who imitated it using a `select-all-that-apply' method. Seeds were assigned a mean of 2.03 emotions per rater (max: 7.11, min: 1.00), with a standard deviation of 1.33 emotions. The proportion of times a given seed sample was rated with each emotion was then applied to all mimics of that seed sample. This method results in the share per emotion assigned by the speakers. The label then is normalized by dividing all (floating point) scores with their highest score per sample; hence, each sample has a maximum score of 1.0 for one emotion category and other scores less than 1.0 for other emotion categories. The baseline results for this dataset is a Spearman score of 0.500 for the development set and 0.514 for the test set (using wav2vec finetuned of affective speech dataset with SVM method). The labels of the test set are not open to the public; hence, only performance on the development/validation data can be calculated directly.

\begin{table}[tbp]
  \centering
  \caption{Summary of the dataset for Emotion Share Recognition, derived from \cite{Schuller2022a}}
  \begin{tabular}{lrrrr}
    \hline & Train & Dev & Test & $\Sigma$ \\
    \hline Sample no. & 30,133 & 12,241 & 9,507 & 51,881 \\
    Sample \%     & 58.1 & 23.6 & 18.3 & 100.0 \\
    Speaker no. & 600 & 202 & 202 & 1,004 \\
    Gender (f:m) & $379: 221$ & $117: 85$ & $141: 61$ & $637: 367$ \\
    Gender \% & $62.9: 37.1$ & $57.9: 42.1$ & $69.8: 30.2$ & $63.5: 36.5$ \\
    \hline
    \end{tabular}
  \label{tab:hpc-request}
\end{table}

Table \ref{tab:hpc-request} shows the characteristics of the dataset. The ratio for training/test split follows the practical machine/deep learning practice, i.e., 80/20. About 18.3\% of the samples are allocated for the test (without labels), while the rest are for training (including validation/development). The ratio of female/male is about 63.5/36.5, with the number of female samples being higher than male. The splitting of training/test follows speaker-independent criteria with a total of 1004 speakers; 20\% of them is allocated for the test set while the rest is for training and development.

\section{Pre-Trained Models}
We evaluated nine pre-trained models as acoustic embedding feature extractors and fused the results of these models for ensemble learning. Ensemble learning leverages the advantages of multiple models to achieve a better performance score. In this study, we evaluated a single late fusion method by averaging the emotion share predictions of different models. The baseline model is the robust version of on wav2vec 2.0 \cite{Baevski2020a} finetuned the affective dataset \cite{Wagner2022a}. For multilingual pre-trained models, we employed XLS-R 53 \cite{Cai2021a} and its variants, XLSR variants \cite{Babu2022a} (XLSR-300M, XLSR-1B, and XLSR-2B). For the XLS-R 53 and XLSR-1B, we also evaluated the finetuned version of this model on the English (EN) and Spanish (SP) datasets. Finetuning for these languages is not available on other XLSR variants. The complete pre-trained models are listed in Table \ref{tab:models}.

\begin{table}[tbp]
  \caption{Pre-trained models used in this study}
  \centering\begin{tabular}{l l}
    \hline
    Model Name & Hugging Face ID/Name \\
    \hline
    wav2vec 2.0 & audeering/wav2vec2-large-robust-12-ft-emotion-msp-dim\\
    XLS-R 53	  & faceboook/wav2vec2-large-xlsr-53 \\
    XLS-R 53 EN & jonatasgrosman/wav2vec2-large-xlsr-53-english \\
    XLS-R 53 SP & jonatasgrosman/wav2vec2-large-xlsr-53-spanish \\
    XLSR-300M	& facebook/wav2vec2-xls-r-300m \\
    XLSR-1B	  & facebook/wav2vec2-xls-r-1b \\
    XLSR-1B EN & jonatasgrosman/wav2vec2-xls-r-1b-english \\
    XLSR-1B SP & jonatasgrosman/wav2vec2-xls-r-1b-spanish \\
    XLSR-2B	    & facebook/wav2vec2-xls-r-2b \\
    \hline
  \end{tabular}
  \label{tab:models}
\end{table}


\section{Classifier and Hyperparameter Search Space}
We employed a support vector machine (SVM) classifier for regression (SVR, support vector regression). The type of kernel for SVR is linear (LinearSVR in the scikit-learn toolkit). The optimal parameters for this SVR were searched using the Grid Search algorithm, i.e., the regularization parameter C and the algorithm to solve either the 'dual' or 'primal' optimization problem. Other parameters like scoring were also searched. The scoring is either using negative mean absolute error (NMAE) or negative mean squared error (NMSE). The maximum iteration is fixed at 5000. The value range for these parameters is shown in Table \ref{tab:params}. 

\begin{table}[htbp]
  \centering
  \caption{Value range for hyperparameter grid search in LinearSVR}
  \begin{tabular}{l l}
    \hline
     Hyperparameter	& Value range \\
    \hline
    Scoring & [NMAE, NMSE] \\
    Scaler  & [StandardScaler, MinMaxScaler] \\
    C & [$10^{-2} - 10^{-6}$] \\
    dual & [True, False] \\
    max\_iter & [50000] \\
    \hline
  \end{tabular}
  \label{tab:params}
\end{table}

For the fusion of all nine models, we employed the average of the predicted values (arithmetic mean). The predicted values are the continuous values in either development or test for each of the nine emotion categories. The ordinal values are then used to calculate performance metrics. The final score is an average of nine emotional categories.

The performance for evaluating the models is measured as the Spearman correlation coefficient ($\rho$) \cite{Spearman1904}. This Spearman's $\rho$ metric is similar to the Pearson correlation coefficient but for ranked data. This metric is chosen since the annotation of the dataset is ordinal (based on rank) and for consistency with the baseline model \cite{Schuller2022a}. Spearman correlation coefficient is calculated as follows:

\begin{equation}
  \rho = \frac{\text{cov}({\text{R}(X),{\text{R}(Y))}}}{\sigma_{\text{R}(X)}\sigma_{\text{R}(Y)}},
\end{equation}
where $\text{cov}(\text{R}(X),\text{R}(Y))$ denotes the covariance of the ranked data $\text{R}(X)$ and $\text{R}(Y)$, and $\sigma_{\text{R}(X)}$ and $\sigma_{\text{R}(Y)}$ denote the standard deviation of the ranked data $\text{R}(X)$ and $\text{R}(Y)$, respectively.

A repository is created to ensure the reproducibility of the experiments \footnote{https://github.com/bagustris/ComParE2023}. The repository contains the experiment settings, including the hyperparameter search space, and the results of the experiments. The requirements to run the experiments are also provided in the repository with the exact version of the libraries used at the time of the experiment.



\section{Experiment Results and Discussions}

We divided our results into validation or development (Table \ref{tab:val-result}) and test results (Table \ref{tab:test-result}). In the development stage, we ensure that the result of ensemble learning (Fusion all in Table \ref{tab:val-result}) is higher than baseline results, either from a single model or late fusion. We submitted our prediction of the test result to the provider of the dataset \cite{Schuller2022a} to obtain the performance on the test set.

Table \ref{tab:val-result} shows our result in the development stage using nine pre-trained models and a fusion of all these models. On a single model evaluation, it is shown that using pre-trained models (such as XLS-R 53 and XLSR variants) achieves higher accuracies than conventional acoustic feature extractors like auDeep and ComParE. For instance, XLS-R 53 (the smallest multilingual model) gained a Spearman score of 0.4328, while ComParE achieved 0.359. For the fusion models, our fusion of all nine pre-trained models also overcomes the previous late fusion of wav2vec2, auDeep, DeepSpectrum, and ComParE. 

\begin{table}[tbp]
  \centering
  \caption{Spearman correlation coefficient ($\rho$) on the development set of the evaluated pre-trained models}
  \begin{tabular}{l l c c}
  \hline
  No. & SSL model & NMSE & NMAE \\
  \hline
  1 & wav2vec 2.0  & 0.5022	& 0.5001 \\
  2 & XLS-R 53	  & 0.4328	& 0.4341 \\
  3 & XLS-R 53 EN	& 0.4732	& 0.4739 \\
  4 & XLS-R 53 SP	& 0.4687	& 0.4691 \\
  5 & XLSR-300M	  & 0.4555	& 0.4561 \\
  6 & XLSR-1B	    & 0.4694	& 0.4679 \\
  7 & XLSR-1B EN 	& 0.4559	& 0.4533 \\
  8 & XLSR-1B SP	& 0.4451	& 0.4440 \\
  9 & XLSR-2B	    & 0.4813	& 0.4835 \\
  & \textbf{Fusion all (of 9)}	& \textbf{0.5236}	& \textbf{0.5266} \\
  \hline
  \end{tabular}
  \label{tab:val-result}
\end{table}

Having good results in the development stage, we evaluated the same model on the test set. Table \ref{tab:test-result} shows the result of the test set. The result shows that the fusion of all models achieves the highest Spearman score of 0.537 (compared to the 0.476 of the late fusion from the previous study \cite{Schuller2022a}). This result is also higher than the previous best result of 0.500 from wav2vec2 \cite{Wagner2022a}. We assume ensemble learning works by leveraging information across different languages and pooling maximum information for multilingual SER.

\begin{table}[tbp]
  \centering
  \caption{Comparison of Spearman correlation coefficient ($\rho$) on the dev(evelopment) and test sets of the baselines (first five rows) and the proposed fusion method (the last row)}
  \begin{tabular}{l l c c}
    \hline
    No. & Model & Dev	& Test \\
    \hline
     1 & wav2vec2   \cite{Wagner2022a}  & 0.500 & 0.514 \\
     2 & auDeep   \cite{Freitag2018}    & 0.347 & 0.357 \\
     3 & DeepSpectrum \cite{amiriparian2017snore} & 0.335 & 0.331 \\
     4 & ComParE \cite{Eyben2013,Schuller2013}     & 0.359 & 0.365 \\
     & Late Fusion (of 4) \cite{Schuller2022a} & 0.470 & 0.476 \\
     \hline
     & \textbf{Fusion all (of 9)} & \textbf{0.524} & \textbf{0.537} \\
    \hline
  \end{tabular}
  \label{tab:test-result}
\end{table}

Similar trends between test and development sets are observed where the performance of the test set is slightly higher than the development set. An exception only applies to the DeepSpectrum result where the test result is 0.004 lower than the development set. This trend indicates the generalization of the evaluated method (including the baseline methods that use a similar SVM classifier) to unseen data and shows that, perhaps, the distribution of the test set is similar to the development set.

Finally, Table \ref{tab:test-emotion} breaks down the average Spearman's $\rho$ in the test set ($\rho=0.537$) into each emotion category. The result is from `Fusion all' of nine pre-trained models on the test set. The result shows that the model performs best in the 'Calmness' emotion category (0.6061) and worst in the `Interest' emotion category (0.4238). The result is similar to the previous study \cite{Schuller2022a} where the `Calmness' emotion category is the highest (0.559), and the `Anger' emotion category is the lowest (0.428). The previous study's result was achieved with wav2vec 2.0 but with a different setup from this study (particularly on the scoring method).

\begin{table}[htbp]
  \caption{Spearman correlation coefficient ($\rho$) on each emotion categories for the test set}
  \centering\begin{tabular}{l c}
    \hline
    Emotion & Spearman's $\rho$ \\
    \hline
    Anger & 0.4894 \\
    Boredom & 0.6032 \\
    Calmness & 0.6061 \\
    Concentration & 0.5855 \\
    Determination & 0.5589 \\
    Excitement & 0.4669 \\
    Interest & 0.4238 \\
    Sadness & 0.5123 \\
    Tiredness & 0.5867 \\
    \hline   
  \end{tabular}
  \label{tab:test-emotion}
\end{table}

\section{Conclusions}
In this study, we evaluated nine pre-trained speech models and a fusion of all these nine models for emotion share recognition from speech. The idea was to collect multilingual models from different languages and apply the fusion of these models to multilingual SER data. A dataset from three countries with English and Spanish languages was selected. First, the fusion of nine pre-trained models defeats the previously reported best result of a single model (in which the performance of this single model is higher than the late fusion of four methods). Second, the results of any single model in this study generally are also higher than the single models/methods in the previous study. Third, there is a trend in which the performance of the test set is slightly higher than the development set. Future studies could be directed to explore more about the dataset and improve the performance using more advanced methods with more recent speech embeddings.

\section*{Acknowledgment}
This paper is partly based on results obtained from a project, JPNP20006, commissioned by the New Energy and Industrial Technology Development Organization (NEDO), Japan.

\bibliography{multiser}

\begin{thebibliography}{10}
\providecommand{\url}[1]{#1}
\csname url@samestyle\endcsname
\providecommand{\newblock}{\relax}
\providecommand{\bibinfo}[2]{#2}
\providecommand{\BIBentrySTDinterwordspacing}{\spaceskip=0pt\relax}
\providecommand{\BIBentryALTinterwordstretchfactor}{4}
\providecommand{\BIBentryALTinterwordspacing}{\spaceskip=\fontdimen2\font plus
\BIBentryALTinterwordstretchfactor\fontdimen3\font minus
  \fontdimen4\font\relax}
\providecommand{\BIBforeignlanguage}[2]{{%
\expandafter\ifx\csname l@#1\endcsname\relax
\typeout{** WARNING: IEEEtran.bst: No hyphenation pattern has been}%
\typeout{** loaded for the language `#1'. Using the pattern for}%
\typeout{** the default language instead.}%
\else
\language=\csname l@#1\endcsname
\fi
#2}}
\providecommand{\BIBdecl}{\relax}
\BIBdecl

\bibitem{Cowen2019}
A.~S. Cowen, H.~A. Elfenbein, P.~Laukka, and D.~Keltner, ``{Mapping 24 emotions
  conveyed by brief human vocalization.}'' \emph{Am. Psychol.}, vol.~74, no.~6,
  pp. 698--712, sep 2019.

\bibitem{Buechel2016}
S.~Buechel and U.~Hahn, ``{Emotion analysis as a regression problem-dimensional
  models and their implications on Emotion representation and metrical
  evaluation},'' \emph{Front. Artif. Intell. Appl.}, vol. 285, pp. 1114--1122,
  2016.

\bibitem{Radford2021}
A.~Radford, J.~Wook, K.~Tao, X.~Greg, B.~Christine, and M.~Ilya, ``{Robust
  Speech Recognition via Large-Scale Weak Supervision},'' \emph{openai.com},
  2021.

\bibitem{Pratap}
V.~Pratap, M.~Auli, and M.~Ai, ``{Scaling Speech Technology to 1 , 000 +
  Languages}.''

\bibitem{Upadhyay2023}
\BIBentryALTinterwordspacing
S.~G. Upadhyay, L.~Martinez-Lucas, B.-h. Su, W.-c. Lin, W.-s. Chien, Y.-t. Wu,
  W.~Katz, C.~Busso, and C.-c. Lee, ``{Phonetic Anchor-Based Transfer Learning
  to Facilitate Unsupervised Cross-Lingual Speech Emotion Recognition},'' in
  \emph{ICASSP 2023 - 2023 IEEE Int. Conf. Acoust. Speech Signal
  Process.}\hskip 1em plus 0.5em minus 0.4em\relax IEEE, jun 2023, pp. 1--5.
  [Online]. Available: \url{https://ieeexplore.ieee.org/document/10095250/}
\BIBentrySTDinterwordspacing

\bibitem{Li2023}
\BIBentryALTinterwordspacing
S.~Li, P.~Song, L.~Ji, Y.~Jin, and W.~Zheng, ``{A Generalized Subspace
  Distribution Adaptation Framework for Cross-Corpus Speech Emotion
  Recognition},'' in \emph{ICASSP 2023 - 2023 IEEE Int. Conf. Acoust. Speech
  Signal Process.}\hskip 1em plus 0.5em minus 0.4em\relax IEEE, jun 2023, pp.
  1--5. [Online]. Available:
  \url{https://ieeexplore.ieee.org/document/10097258/}
\BIBentrySTDinterwordspacing

\bibitem{Zhao}
Y.~Zhao, J.~Wang, Y.~Zong, W.~Zheng, H.~Lian, and L.~Zhao, ``{DEEP IMPLICIT
  DISTRIBUTION ALIGNMENT NETWORKS FOR CROSS-CORPUS SPEECH EMOTION
  RECOGNITION},'' \emph{ICASSP 2023 - 2023 IEEE Int. Conf. Acoust. Speech
  Signal Process.}

\bibitem{Atmaja2020h}
B.~T. Atmaja, Y.~Hamada, and M.~Akagi, ``{Predicting Valence and Arousal by
  Aggregating Acoustic Features for Acoustic-Linguistic Information Fusion},''
  in \emph{2020 IEEE Reg. 10 Conf.}\hskip 1em plus 0.5em minus 0.4em\relax
  IEEE, nov 2020, pp. 1081--1085.

\bibitem{Atmaja2020e}
B.~T. Atmaja and M.~Akagi, ``{Improving Valence Prediction in Dimensional
  Speech Emotion Recognition Using Linguistic Information},'' in \emph{Proc.
  2020 23rd Conf. Orient. COCOSDA Int. Comm. Co-ord. Stand. Speech Databases
  Assess. Tech. O-COCOSDA 2020}.\hskip 1em plus 0.5em minus 0.4em\relax IEEE,
  nov 2020, pp. 166--171.

\bibitem{Zhao2018}
J.~Zhao and S.~Chen, ``{Multi-modal Multi-cultural Dimensional Continues
  Emotion Recognition in Dyadic Interactions},'' in \emph{Cross-cultural Emot.
  Sub-challenge AVEC'18}, 2018, pp. 65--72.

\bibitem{Atmaja2020f}
B.~T. Atmaja and M.~Akagi, ``{The Effect of Silence Feature in Dimensional
  Speech Emotion Recognition},'' in \emph{10th Int. Conf. Speech Prosody 2020},
  no. May.\hskip 1em plus 0.5em minus 0.4em\relax ISCA, may 2020, pp. 26--30.

\bibitem{Shih2017}
P.-Y. Shih, C.-P. Chen, and C.-H. Wu, ``{Speech Emotion Recognition With
  Ensemble Learning Methods},'' in \emph{IEEE Int. Conf. Acoust. Speech, Signal
  Process. 2017}, 2017, pp. 2756--2760.

\bibitem{Schuller2022a}
\BIBentryALTinterwordspacing
B.~Schuller, A.~Batliner, S.~Amiriparian, C.~Bergler, M.~Gerczuk, N.~Holz,
  P.~Larrouy-Maestri, S.~Bayerl, K.~Riedhammer, A.~Mallol-Ragolta, M.~Pateraki,
  H.~Coppock, I.~Kiskin, M.~Sinka, and S.~Roberts, ``{The ACM Multimedia 2023
  Computational Paralinguistics Challenge: Emotion Share and Requests},'' in
  \emph{Proc. 30th ACM Int. Conf. Multimed.}\hskip 1em plus 0.5em minus
  0.4em\relax New York, NY, USA: ACM, oct 2023, pp. 7120--7124. [Online].
  Available: \url{https://dl.acm.org/doi/10.1145/3503161.3551591}
\BIBentrySTDinterwordspacing

\bibitem{Cowen2019a}
\BIBentryALTinterwordspacing
A.~S. Cowen, P.~Laukka, H.~A. Elfenbein, R.~Liu, and D.~Keltner, ``{The primacy
  of categories in the recognition of 12 emotions in speech prosody across two
  cultures},'' \emph{Nat. Hum. Behav.}, vol.~3, no.~4, pp. 369--382, apr 2019.
  [Online]. Available: \url{http://www.nature.com/articles/s41562-019-0533-6}
\BIBentrySTDinterwordspacing

\bibitem{Poria2019}
S.~Poria, D.~Hazarika, N.~Majumder, G.~Naik, E.~Cambria, and R.~Mihalcea,
  ``{MELD: A Multimodal Multi-Party Dataset for Emotion Recognition in
  Conversations},'' in \emph{Proc. 57th Annu. Meet. Assoc. Comput. Linguist.},
  2019, pp. 527--536.

\bibitem{Laukka2010}
P.~Laukka, H.~A. Elfenbein, W.~Chui, N.~S. Thingujam, F.~K. Iraki,
  T.~Rockstuhl, and J.~Althoff, ``{Presenting the VENEC Corpus: Development of
  a Cross-Cultural Corpus of Vocal Emotion Expressions and a Novel Method of
  Annotating Emotion Appraisals},'' \emph{Lr. 2010 - Seventh Int. Conf. Lang.
  Resour. Eval.}, vol.~7, pp. 53--57, 2010.

\bibitem{Baevski2020a}
A.~Baevski, H.~Zhou, A.~Mohamed, and M.~Auli, ``{wav2vec 2.0: A framework for
  self-supervised learning of speech representations},'' \emph{Adv. Neural Inf.
  Process. Syst.}, 2020.

\bibitem{Wagner2022a}
J.~Wagner, A.~Triantafyllopoulos, H.~Wierstorf, M.~Schmitt, F.~Burkhardt,
  F.~Eyben, and B.~W. Schuller, ``{Dawn of the Transformer Era in Speech
  Emotion Recognition: Closing the Valence Gap},'' \emph{IEEE Trans. Pattern
  Anal. Mach. Intell.}, pp. 1--13, mar 2023.

\bibitem{Cai2021a}
X.~Cai, Z.~Wu, K.~Zhong, B.~Su, D.~Dai, and H.~Meng, ``{Unsupervised
  Cross-Lingual Speech Emotion Recognition Using Domain Adversarial Neural
  Network},'' in \emph{2021 12th Int. Symp. Chinese Spok. Lang. Process. ISCSLP
  2021}, 2021, pp. 3--7.

\bibitem{Babu2022a}
\BIBentryALTinterwordspacing
A.~Babu, C.~Wang, A.~Tjandra, K.~Lakhotia, Q.~Xu, N.~Goyal, K.~Singh, P.~von
  Platen, Y.~Saraf, J.~Pino, A.~Baevski, A.~Conneau, and M.~Auli, ``{XLS-R:
  Self-supervised Cross-lingual Speech Representation Learning at Scale},'' in
  \emph{Interspeech 2022}, vol. 2022-Septe.\hskip 1em plus 0.5em minus
  0.4em\relax ISCA: ISCA, sep 2022, pp. 2278--2282. [Online]. Available:
  \url{https://www.isca-speech.org/archive/interspeech{\_}2022/babu22{\_}interspeech.html}
\BIBentrySTDinterwordspacing

\bibitem{Spearman1904}
\BIBentryALTinterwordspacing
C.~Spearman, ``{The Proof and Measurement of Association between Two Things},''
  \emph{Am. J. Psychol.}, vol. 100, no. 3/4, p. 441, 1987. [Online]. Available:
  \url{https://www.jstor.org/stable/1422689?origin=crossref}
\BIBentrySTDinterwordspacing

\bibitem{Freitag2018}
M.~Freitag, S.~Amiriparian, S.~Pugachevskiy, N.~Cummins, and B.~Schuller,
  ``{auDeep: Unsupervised learning of representations from audio with deep
  recurrent neural networks},'' \emph{J. Mach. Learn. Res.}, vol.~18, pp. 1--5,
  2018.

\bibitem{amiriparian2017snore}
S.~Amiriparian, M.~Gerczuk, S.~Ottl, N.~Cummins, M.~Freitag, S.~Pugachevskiy,
  A.~Baird, and B.~Schuller, ``{Snore Sound Classification Using Image-Based
  Deep Spectrum Features},'' in \emph{Interspeech 2017}.\hskip 1em plus 0.5em
  minus 0.4em\relax ISCA: ISCA, aug 2017, pp. 3512--3516.

\bibitem{Eyben2013}
F.~Eyben, F.~Weninger, F.~Gross, and B.~Schuller, ``{Recent developments in
  openSMILE, the munich open-source multimedia feature extractor},'' in
  \emph{Proc. 21st ACM Int. Conf. Multimed. - MM '13}.\hskip 1em plus 0.5em
  minus 0.4em\relax New York, New York, USA: ACM Press, 2013, pp. 835--838.

\bibitem{Schuller2013}
B.~Schuller, S.~Steidl, A.~Batliner, A.~Vinciarelli, K.~Scherer, F.~Ringeval,
  M.~Chetouani, F.~Weninger, F.~Eyben, E.~Marchi, M.~Mortillaro, H.~Salamin,
  A.~Polychroniou, F.~Valente, and S.~Kim, ``{The INTERSPEECH 2013
  computational paralinguistics challenge: social signals, conflict, emotion,
  autism},'' in \emph{Interspeech 2013}, no. August.\hskip 1em plus 0.5em minus
  0.4em\relax ISCA: ISCA, aug 2013, pp. 148--152.

\end{thebibliography}
\bibliographystyle{IEEEtran}
\end{document}